\documentstyle[12pt]{article}
\newcommand{\vect}[1]{\mbox{\bf #1}}

\newcommand{\rms}[1]{\mbox{\scriptsize #1}}

\newcommand{\lw}[1]{\smash{\lower 1.5ex\hbox{#1}}}

\begin{document}
%%%%%%%%%%%%%%%%%%%%%%%%%%%%
% \begin{flushright}
%    hep-ph/9606365
% \end{flushright}
%
%   TITLE & AUTHORS
%
%%%%%%%%%%%%%%%%%%%%%%%%%%%%%%
\title{\Large\bf
 Coulomb and strong interactions for Bose-Einstein correlations
%       \vspace{-3mm}\\
 \vspace{-2mm}\\}
\author{T. Osada$^1$,
%\thanks{e-mail: minoru44@jpnyitp.bitnet},
         S. Sano$^2$  and M. Biyajima$^2$\\
{\small $^1$Department of Physics, Tohoku University, Sendai 980,
Japan}\\
{\small $^2$Department of Physics, Faculty of Science,}\\
%\vspace{-4mm}\\
{\small Shinshu University, Matsumoto 390, Japan}\\}
\date{\today}
%\date{}

\maketitle
\begin{abstract}
We present an analytical formula for the Bose-Einstein
correlations (BEC) which includes effects of both Coulomb
and strong final state
interactions (FSI). It was obtained by using Coulomb wave function
together with the scattering partial wave amplitude of the strong
interactions describing data on the $s$-wave phase shift. We have
proved numerically that this method is equivalent to solving
Schr\"{o}dinger equation with Coulomb and the $s$-wave strong
interaction potentials. As an application we have analysed,
using our formula which includes the degree of coherence and the
long range correlation, the data for $e^+e^-$ annihilations. 

We have found that the degree of coherence present in our
formula approaches approximately unity whereas the long range
correlation parameter becomes approximately zero. These results
suggest that the physical meanings of the fractional degree of
coherence and the long range correlation observed in various other
analyses can most probably be attributed to FSI.
\end{abstract}
%
%  Section Introduction
%
\section {Introduction}
Very recently several analytical expressions for the final state
interaction (FSI) corrections to Bose-Einstein correlations (BEC)
due to Coulomb interaction have been presented
in \cite{biya95,biya95J}
(see also \cite{bowler91,Lednicky}). On the other hand it is also
well known that there exists FSI due to the strong interactions
between identical pions \cite{suzuki87,bowler88,biya95zp}. In order
to treat both the FSI of the Coulomb type and those emerging from
the strong interactions at the same time, one has to solve
Schr\"{o}dinger equation with the corresponding Coulomb and strong
potentials. Such study has been made by Pratt et al. \cite{pcz90}
\footnote{Study exploiting WKB method is presented
in \cite{anzi95}. See also \cite{koonin} and \cite{biya83},
where $pp$ correlations were studied.}. However,
in approach of \cite{pcz90} it is
difficult to elucidate the physical meaning of the degree of
coherence parameter introduced in various analyses of data on
the BEC (see below).\\

In this paper we consider this problem using different approach from
that of \cite{pcz90}. Our aim will be to obtain an analytical formula
for the BEC which would include also FSI caused by both Coulomb and
strong interactions. Comparing our results with those of \cite{pcz90}
we prove numerically that both methods are equivalent. The data of
the BEC in $e^{+}e^{-}$annihilation by TPC, AMY, OPAL ALEPH and
DELPHI Collaborations \cite{aihara85,amy95,opal93,aleph92,delphi92}
are then analysed using our formula. Finally concluding remarks are
presented.\\
%
% SECTION WAVE FUNCTION
%
\section {Phase shift and wave function of identical
$\pi$-$\pi$ scattering}
Let us first consider the Coulomb wave function
of the identical $\pi$-$\pi$ scattering with momenta
$p_1$ and $p_2$
\cite{schiff,sasagawa}:
\begin{eqnarray}
% \Psi_{total}(\bok,\bor) &=& \Psi_C(\bok,\bor) +
 %\phi_{st}(\bok,\bor), \\
  \Psi_{\rms{C}}(\vect{k},\vect{r})
  &=& \Gamma(1+i\eta) e^{-\pi\eta/2}
             e^{i {\vect{\small{k}} \cdot \vect{\small{r}}}}
             F(-i\eta;1;ikr(1-\cos\theta)), \\
        &=& \sum_{l=0}^{\infty} (2l+1) i^l e^{i\eta_l}
             R_l(kr) P_l(\cos\theta). \nonumber
\end{eqnarray}
Here $F$ denotes the confluent hypergeometric function and
\begin{eqnarray}
  2k &=& Q = p_1 - p_2, \nonumber \\
  \eta &=& \frac{m \alpha}{Q}, \nonumber \\
  \eta_l &=& \arg \Gamma(l+1+i\eta), \nonumber \\
  R_l(kr) &=& e^{-\pi\eta/2} \frac{|\Gamma(l+1+i\eta)|}{(2l+1)!}
     (2kr)^l e^{ikr} F(l+1+i\eta;2l+2;-2ikr). \nonumber
\end{eqnarray}
%The asymptotic formula of the Coulomb wave
%function is well known and
%given by: \\

In numerical calculations of the hypergeometric function appearing in
(1) wild (oscillating) behavior emerges when $kr$ becomes large.
To avoid the computational problem in this asymptotic region, we use
there the following asymptotic formula of the Coulomb wave function:
\begin{eqnarray}
  \Psi_{\rms{C}}^{\rms{asym}} (\vect{k}, \vect{r})
     &=& \exp \{ i(kz + \eta\ln(k(r-z))) \}
    \left[ 1 + \frac{\eta^2}{ik(r-z)} \right] \nonumber \\
     & & + f(\theta) \frac{\exp\{i(kr - \eta \ln(2kr))\}}r
         \:,\label{eq:asym1}
\end{eqnarray}
where $z = r \cos \theta$ and the scattering amplitude
$f(\theta)$ is given as follows:
\begin{eqnarray*}
  f(\theta) = - \frac{\eta}{2k} \frac 1{\sin^2 (\theta/2)}
   \exp\{ -2i\eta\ln \sin (\theta/2) + 2i \arg
\Gamma (1+i\eta) \}.
\end{eqnarray*}
The asymptotic formula for an $s$-wave component
of the Coulomb wave
function is obtained in the similar way \cite{sasagawa}:
\begin{eqnarray}
  \Psi_{\rms{C(s-wave)}}^{\rms{asym}} (\vect{k}, \vect{r})
 &=&   \exp \{ i(kr - \eta\ln(2kr) + 2\eta_0) \} \frac{1}{2ikr}
 \left[ 1 + \frac{i\eta(1+i\eta)}{2ikr} \right] \nonumber \\
   & & + \exp \{-i(kr - \eta\ln(2kr)) \} \frac{1}{-2ikr}
           \left[ 1 + \frac{i\eta(1-i\eta)}{2ikr} \right].
\end{eqnarray}

As the next step let us consider the strong interaction provided in
terms of the phase shift in the identical $\pi$-$\pi$ scattering. The
data for the $s$-wave ($I$ = 2) phase shift reported in
\cite{walker70} -\cite{hoog74} can be described phenomenologically by
the following formula \cite{suzuki87}:
\begin{equation}
 \delta_{0}^{(2)}
     = \frac{1}{2}
      \left( \frac{a_0Q}{1+0.5Q^2} \right),  \label{phas}
\end{equation}
where parameter $a_0$  ( -1.5 $\le a_0 \le$ -0.7 (GeV${}^{-1})$~)
denotes the scattering length.\\

Finally, using (4), the scattering $s$-wave function induced by
strong interaction is expressed as:
\begin{eqnarray}
  \phi_{\rms{st}}(\vect{k}, \vect{r})
      =f^0(\theta) \frac{\exp\{i(kr - \eta \ln(2kr)
)\}}r,\label{strasy} \\
  f^0(\theta)
    = \frac{1}{2ik} \exp(2i\eta_0) (\exp
(2i\delta_{0}^{(2)})-1).\nonumber
\end{eqnarray}
%
%Combining (4) and (5) we obtain therefore the final form for the
%wave function $\phi_{st}$.\\
%It should be noticed here (cf. Fig. 1(a)) that we need two wave
%functions of the scattering of identical bosons ($\pi$'s)
%defined in
%two regions: the region ruled by the strong interactions and the
%asymptotic region.
It should be noticed here (cf. Fig. 1(a)) that in addition to (5) we
need one more $\pi$-$\pi$ scattering wave function, which would
reasonably describe the strong interaction in the small $kr$ region.
%%%%%%%%%%%%
%because the exact wave function of the strong interaction
%is not known in that region.
%\footnote{
%We take this region same as the non-asymptotic region
%of the Coulomb interaction.}.
%%%%%%%%%%%%%
Of course Coulomb potential affects both regions, cf. an "unknown
box" in Fig.1(a). To obtain a wave function suitable for that box we
use the following assumption proposed by Bowler \cite{bowler88}
\footnote{Cf. also \cite{biya95zp} where the plane wave and the data
of the s-wave phase shift of identical $\pi$-$\pi$ scattering were
used. In the present calculation we have to use the Coulomb wave
function instead of the plane wave.}: We assume that the wave
function given by (5) with a renormalization provided by the square
root of the Gamow factor can be interpolated into the internal
region. This is attributed to the normalization of the Coulomb wave
function which is given by the following factor:
$\sqrt{G(k)} = (2\pi\eta/(\exp(2\pi\eta)-1))^{1/2}$.\\

For the asymptotic region we use the following expression:
\begin{equation}
\Psi_{\rms{total}}(\vect{k},\vect{r}) =
\Psi_{\rms{C}}^{\rms{asym}}(\vect{k},\vect{r})
+ \phi_{\rms{st}}(\vect{k},\vect{r})
\end{equation}
whereas for the region described by the exact Coulomb wave
function the following wave function is used instead:
\begin{equation}
\Psi_{\rms{total}}(\vect{k},\vect{r}) =
\Psi_{\rms{C}}(\vect{k},\vect{r})
+ \sqrt{G(k)}\phi_{\rms{st}}(\vect{k},\vect{r}).
\end{equation}
As seen in Fig.1(b), there is smooth connection between
both regions. The usefulness of this assumption will show
up in Section 4.
%
% SECTION FORMULATION
%
%
 \section  {Formulation of BEC}

To describe a pair of the identical bosons, we have to
symmetrize the total wave function in the following way:
\begin{equation}
  A_{12} = \frac{1}{\sqrt{2}}
 [ \Psi_{\rms{C}}(\vect{k},\vect{r})
+ \Psi_{\rms{C}}^{S}(\vect{k},\vect{r})
 + \Phi_{\rms{st}}(\vect{k},\vect{r})
+ \Phi_{\rms{st}}^{S}(\vect{k},\vect{r})],
\end{equation}
where superscript $S$ denotes the symmetrization of the wave
function. The function $ \Phi_{\rms{st}}(\vect{k},\vect{r}) $ stands
for the wave function induced by strong interactions. Assuming a
source function $\rho(r)$ we obtain the following expression for the
BEC including the FSI:
\begin{eqnarray}
   N^{(\pm\pm)}/N^{BG} &=& \frac{1}{G(k)} \int_{}^{}
   \rho(r) d^3r|A_{12}|^2,
                  \nonumber \\
   &=& I_{\rms{C}} + I_{\rms{Cst}} + I_{\rms{st}}, \\
   I_{{\rms C}}   &=& \sum_{m,n=0}^{\infty}
               \frac{1}{m+n+1}
       I_{R1}(2+m+n) A_1(n)A_1^{\ast}(m) \nonumber \\
           & & \times
               \left[
                      1+\frac{n!m!}{(n+m)!}
                     \left( 1+\frac{n}{i\eta} \right)
                     \left( 1-\frac{m}{i\eta} \right)
               \right] \\
     &=& (1 + \Delta_{\rms{1C}}) + (E_{2\rms{B}} +
\Delta_{\rms{EC}}), \\
   I_{{\rms Cst}} &=& 2\Re
               \left[
        \frac{2}{k} (2k)^{i\eta}
   \exp{(-i(\eta_0+\delta_{0}^{(2)}))} \sin\delta_{0}^{(2)}
    \sum_{n=0}^{\infty} I_{\rms{R2}}(1+n) A_2(n,0)
               \right], \\
   I_{{\rms st}}  &=& \frac{2}{k^2}
I_{\rms{R1}}(0) \sin^2\delta_{0}^{(2)},
\end{eqnarray}
where
\begin{eqnarray*}
%   E_{2B} &=& \exp(-\beta^2 Q^2/2), \\
   E_{2{\rms B}} &=& \int_{}^{} d^3r~
\rho(r) ~
   {\textstyle e}^{-{\rms i}
   {\scriptstyle {\bf Q}} \cdot
   {\scriptstyle {\bf r}} }, \\
   1 + \Delta_{1{\rms C}} &=& 1+4\pi \cdot 2\eta
                          \int_{}^{}\rho r^2dr
     \sum_{n=0}^{\infty}\frac{(-1)^n A^{2n+1}}{(2n+1)!
(2n+1)(2n+2)}, \\
   A_1(n)    &=& \frac{\Gamma(n+i\eta)}{\Gamma(i\eta)}
                 \frac{(-2ik)^n}{(n!)^2}, \\
   A_2(n,l)  &=& \frac{\Gamma(n+l+1+i\eta)\Gamma(2l+2)}
                      {\Gamma(l+1+i\eta)\Gamma(n+2l+2)}
                 \frac{(-2ik)^n}{n!}, \\
   I_{{\rms R1}}(n) &=& 4\pi \int_{}^{}dr r^n \rho(r), \\
   I_{{\rms R2}}(n) &=& 4\pi \int_{}^{}dr r^{n+i\eta} \rho(r).
\end{eqnarray*}
In this paper $ N^{(\pm\pm)}/N^{BG} $ stands for the ratio of pairs
of identical charged bosons in a single event to those from different
events. Whenever the data are corrected by the Gamow factor the final
formula should also be divided by the Gamow factor. In numerical
computations we have to assume explicitly some forms of the source
function. In present calculation we use the  gaussian source function:
$\rho(r)=(\frac{1}{\sqrt{2 \pi } \beta})^3
\exp (\frac{-r^2}{2 \beta^2})$.
Therefore the Fourier transform of the source function is given as
follows:
\begin{eqnarray*}
E_{2 {\rms B}} &=& \exp(-\beta^2 Q^2/2).\\ \nonumber
\end{eqnarray*}
%
% SECTION  COMPARISONS OF OUR RESULTS
%              WIDTH THOSE OF PRATT ET AL.
%
\section {Comparisons of our results with those of Pratt et al.}
Authors of \cite{pcz90} have presented their results for the BEC with
interaction ranges $\beta = 2$ fm and $ \beta  = 20$ fm
(corresponding to $R $ in their notation) by solving the following
Schr\"{o}dinger equation:
\begin{equation}
  \left[
        \frac{d^2}{dx^2} + K^2 - \frac{L(L +1)}{x^2}
      - \frac{\epsilon}{x} - U(x)
  \right] \psi_L^I(x)
  = 0, \nonumber\\
\end{equation}
where $\mu = m/2$, $x = \mu r$, $K^2 = 2E / \mu$,
$\epsilon =2\alpha$ and $U(x) =
(2V_0/m_{\rho} x) \exp(-m_{\rho} x / \mu)$ for the $s$-wave. The
values of parameters: $V_0 =  2.6$ GeV and $m_{\rho}= 0.77 $ GeV
were used. In Fig. 2 we compare our results  with theirs using the
same values of parameters $(\beta  $ = 2 fm and $ \beta $ = 20 fm).
Because $I_{\rms{C}} $ with $\beta = 20 $ fm shows wild oscillations
near 50 MeV/c, which are due to the series expansion of the confluent
hypergeometric function $(n+m=75)$, in our calculation we have used a 
method of seamless fitting introduced in \cite{biya95J}. If we use a
different constraint, $n+m=50$, we observe a sharp decreasing near 42
MeV/c, all this depends on the parameter $\beta$. The origin of this
phenomenon can be attributed to the mathematical property of the
confluent hypergeometric functions and the ability of computers.
Therefore in the asymptotic region (2) should be used. \\

As seen in Fig. 2 our results and solutions of the Schr\"{o}dinger
equation \cite{pcz90} are numerically equivalent to each other except
for the behavior near $Q \cong $ a few MeV which is due to the
logarithmic term in (5). It is therefore confirmed that from
their potential ($V_0$ = 2.6 GeV) we obtain in the Born approximation
the scattering length $a_0$ = - 0.8$~\sim$~- 0.6 GeV${}^{-1}$. \\

In the actual analyses we have to introduce a cutoff parameter in the
small $Q$ region. Since in many cases there are no data, or available
data have large error bars due to limits of momentum resolutions,
analyses of data do not critically depend of this cutoff parameter.
%
% SECTION  ANALYSES OF DATA BY TPC,AMY,OPAL,
%                     ALEPH AND DELPHI COLLABORATIONS
%
\section {Analyses of data in $e^+e^-$ annihilation}
As stressed in Sec.1, we want to elucidate the physical meaning of
the degree of coherence. The parameter $\lambda$ describing it should
be therefore introduced into (9) in the usual way. Moreover, notice
that two more parameters: the additional normalization factor $c$ and
the long range correlation parameter $\gamma$ are also introduced by
hand. Our final formula containing all these parameters is thus given
as:
\begin{eqnarray}
N^{(\pm\pm)}/N^{\rms{BG}}
(Q = 2k) &=& c( 1 + \Delta_{\rms{1C}}
  + \Delta_{\rms{EC}}+ I_{\rms{Cst}} + I_{\rms{st}}) \nonumber \\
  & & \times
      \left[
           1 +
     \lambda \frac{ E_{\rms{2B}} }
    { 1 + \Delta_{\rms{1C}}+\Delta_{\rms{EC}}+
I_{\rms{Cst}} + I_{\rms{st}} }
      \right]
      (1 + \gamma Q).
      \hspace{1cm} \label{eq:ratio}
\end{eqnarray}
It should be noted that the normalization $c$ and an effective
degree of coherence, i.e., the denominator of the ratio
$E_{\rms{2B}}/( 1 + \Delta_{\rms{1C}} + \Delta_{\rms{EC}} +
I_{\rms{Cst}} + I_{\rms{st}})$, are related to each other.\\

For the sake of reference we use in our analyses also the
conventional formula (i.e., the standard formula without
corrections due to the FSI):
\begin{equation}
  N^{(\pm\pm:\rms{Standard})}/N^{\rms{BG}} (Q = 2k)
         = c\left[1 + \lambda E_{\rms{2B}} \right]
           (1 + \gamma Q).
%          \label{eq:a22}
\end{equation}
We apply our formulae to data for $e^+e^-$ annihilation provided by
\cite{aihara85}-\cite{delphi92}. Results of our analyses performed by
means of (15) and (16) are shown in Fig. 3 and Table I.
\footnote{It is difficult to treat $a_0$ as a free parameter in the
CERN MINUIT program, due to the limited ability of our computer.}  As
seen in Table I, estimated values of the degree of coherence
parameter $\lambda$ are systematically larger (approaching unity)
when (15) is used than those obtained by the standard formula (16).
On the other hand, estimated values of the long range correlation
parameter $\gamma$ approach approximately zero when (15) is used
(except for the result found in present analysis of data by AMY
collaboration).\\
%We also found that,
%the size parameter $\beta$ estimated by using (15) is
%larger than value
%reported by the OPAL,~ALEPH and DELPHI collaboration at LEP.
%On the other hand, for below the LEP energy,
%$\beta$ estimated by using (15) is smaller
%than the value reported
%PEP4-~TPC and AMY collaboration.\\

Finally please notice that reported values in \cite{aihara85}
-\cite{delphi92} are obtained by the following formula:
\begin{equation}
  N^{(\pm\pm:\rms{Empirical.})}/N^{\rms{BG}} (Q = 2k)
         = c\left[1 + \lambda \exp(-Q^2R^2) \right]
           (1 + \gamma Q).
%          \label{eq:a22}
\end{equation}
To compare our estimated values with various reported ones, we have
to use a relation,  $ \beta = \sqrt{2}R $, because of the different
method of integrations. In Table I we show therefore the corrected
values instead of the values reported by various collaborations.
%%%%%%%%%%%%%%%%%%
%When the correlation function is parametrized
%by using the functional form: $1+\lambda \exp( -Q^2R^2)$,
%our source size parameter $\beta$ should be compared
%with $\sqrt{2}R$.
%For the $e^+e^-$ annihilation data obtained by
%PEP4-~TPC, AMY, OPAL, ALEPH, and DELPHI collaboration,
%   as compared with the value reported by those experimental
%  group and
% obtained by using the standard formula (16).
%%%%%%%%%%%%%%%%%%
The $\beta$'s obtained in the analysis by (15) are systematically
smaller than the corrected values taken from \cite{aihara85}
-\cite{delphi92} and the estimated ones obtained in the analysis
by the standard formula (16).

%
% SECTION CONCLUDING REMARKS
%
\section{Concluding remarks}
We obtain analytic formula (9) for BEC including the Coulombic
and strong FSI. It is (numerically) confirmed that our method is
equivalent to solving the Schr\"{o}dinger equation (14). \\

Combining the seamless fitting method \cite{biya95J} and the CERN
MINUIT program in (15) we have analysed data for BEC in $e^{+}e^{-}$
annihilations. Our results are significantly different from those
obtained by the standard formula (16). Namely, it is found that the
degree of coherence parameter $\lambda$ and the long range
correlation  parameter $\gamma$ approaches approximately unity and
zero, respectively (see Table I). Therefore, we conclude that the
physical meanings of the fractional degree of coherence parameter
$\lambda$ and the long range correlation parameter $\gamma$ as
obtained by the standard formula (16) should be attributed to the
FSI. Moreover, the values of the source size parameter reported by
various collaborations (after using the relation: $\beta =
\sqrt{2}R$) and obtained by the standard formula (16) are
systematically larger than values estimated by (15).

\vspace{1cm}

{\bf Acknowledgements:}~~~The authors would like to thank T.
Mizoguchi, H. Sagawa, T. Sasakawa, F. Shibata and T. Ueda for their
kind correspondences. Numerical computations are partially done by the
computer at Bubble Chamber physics Laboratory (Tohoku). This work is
partially supported by Japanese Grant-in-Aid for Scientific Research
>from the Ministry of Education, Science, Sport and Culture (\#
06640383). Finally the authors are also indebted to G. Wilk for his
reading the manuscript.
%%%%%%%%%%

%%%%%%%%%%%%%%%%%%%%%%%%%%
\newpage
\begin{center}
Figure Captions\\
\end{center}

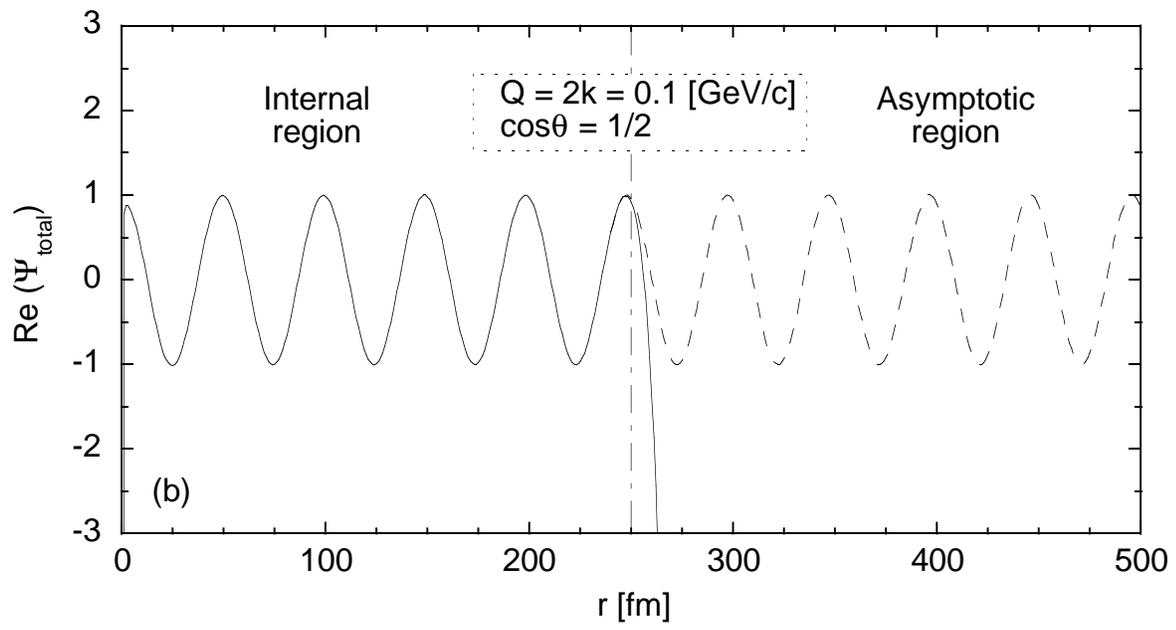
Fig. 1. (a) Interrelation between the two wave functions in
external and internal regions.  (b) Real part of total wave
function with $Q$ = 100 MeV/c and $\theta = \pi/2 $. See (7).
Near 250 MeV/c  there is a connecting point.\\

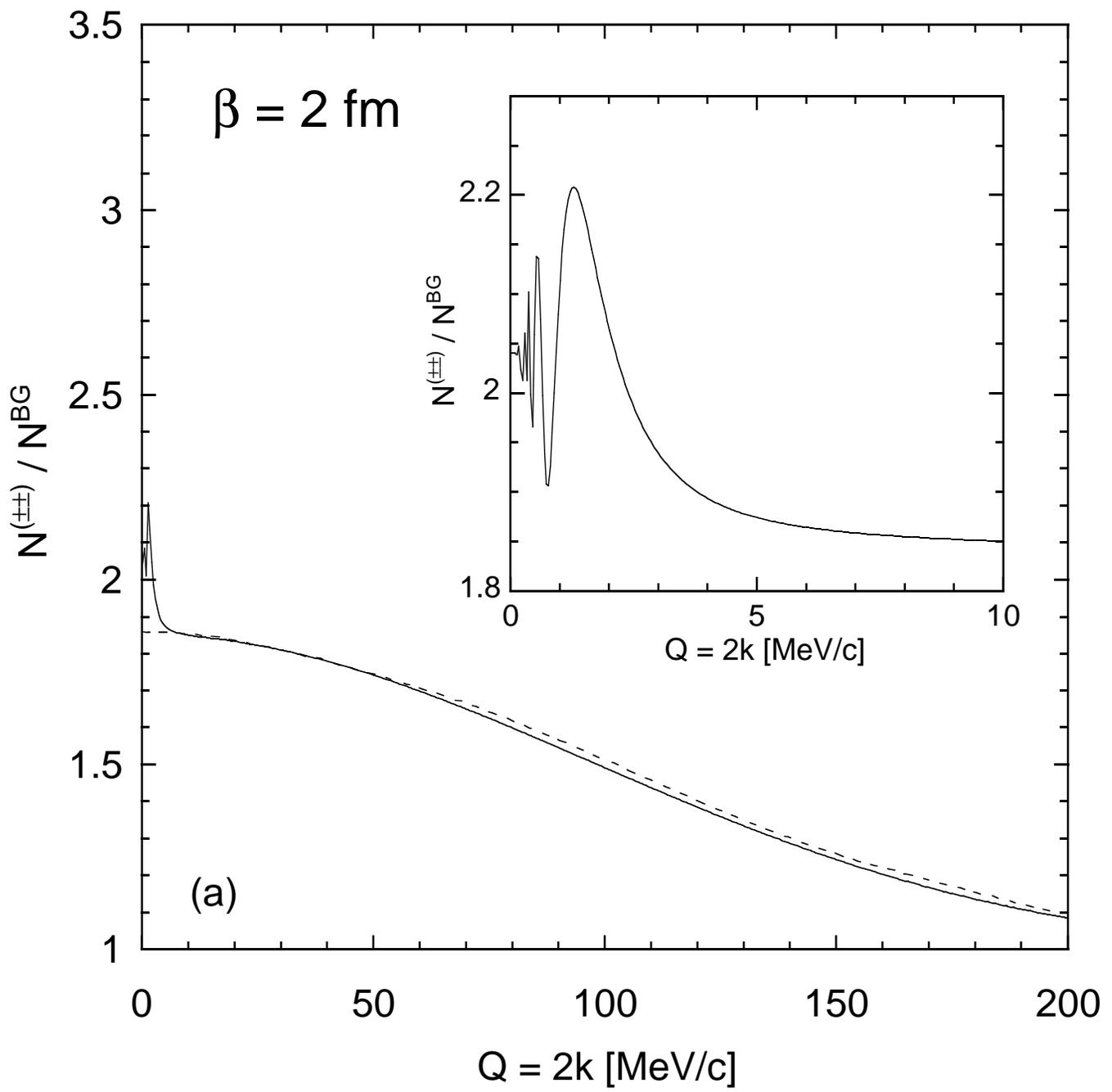
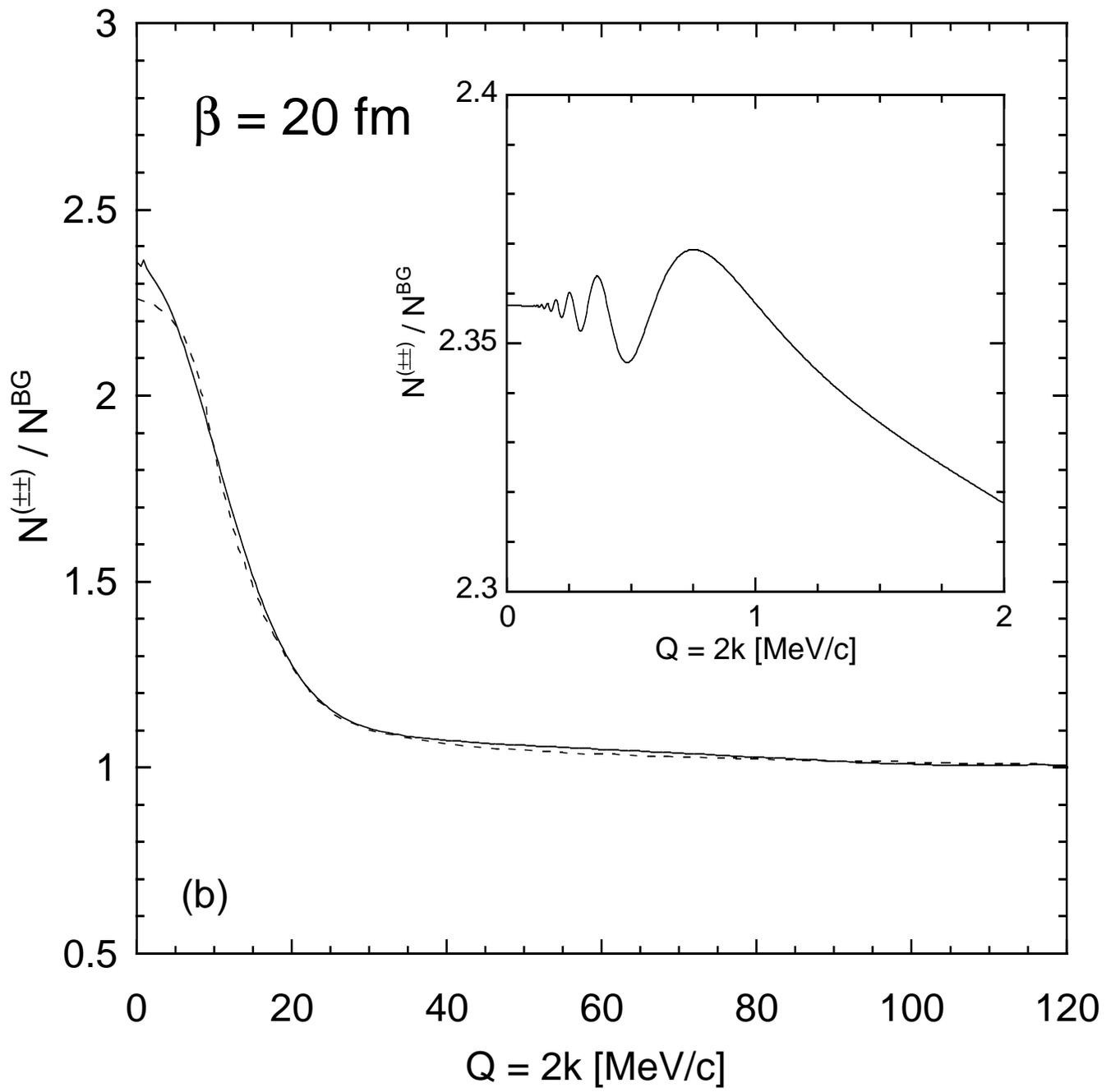
Fig. 2.  Comparisons of our results (solid line) and
those of Pratt et al., \cite{pcz90} (dashed line).
The scattering length $a_0 = -0.6
$ GeV${}^{-1}$  is used. This value approximately
corresponds to $V_0
$ = 2.6 GeV in the Born approximation. (a) $\beta $ = 2 fm
and (b)
$\beta $ = 20 fm. Behaviors in small $Q$ region
(see inserts) are
attributed to the logarithmic term in (5).\\

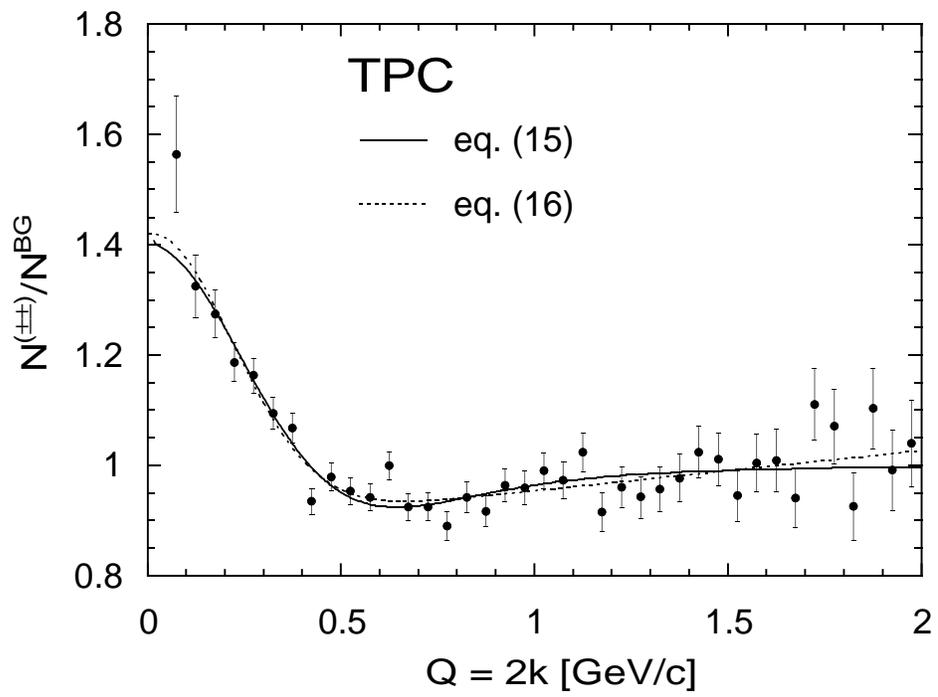
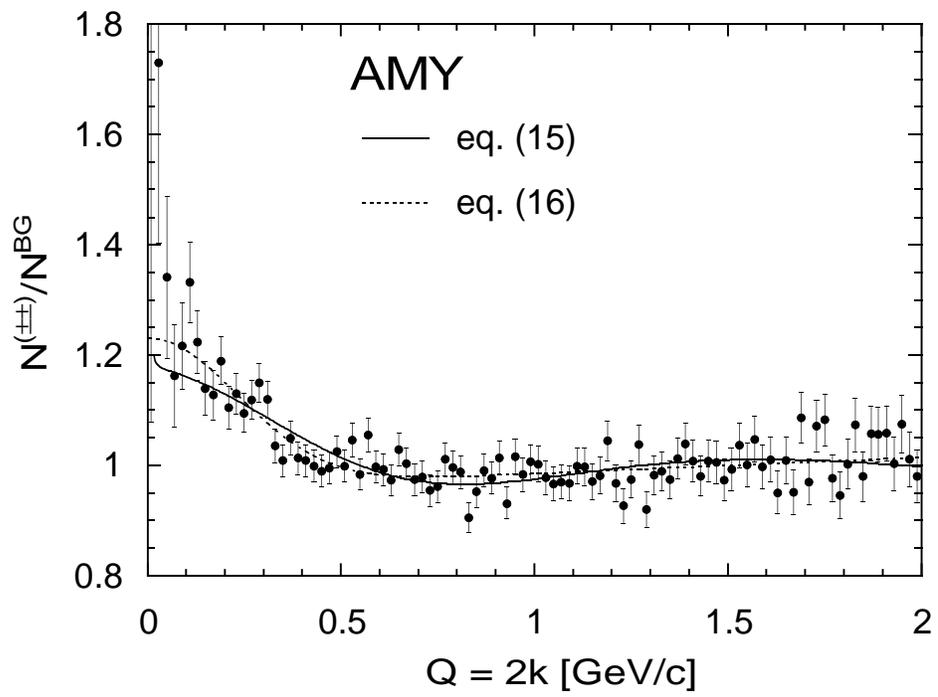
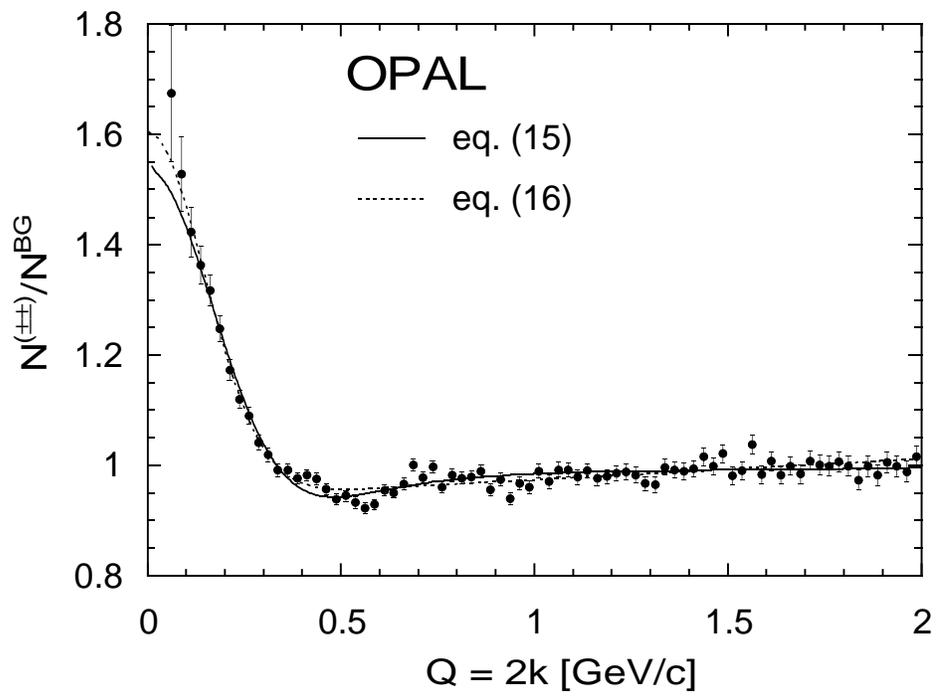
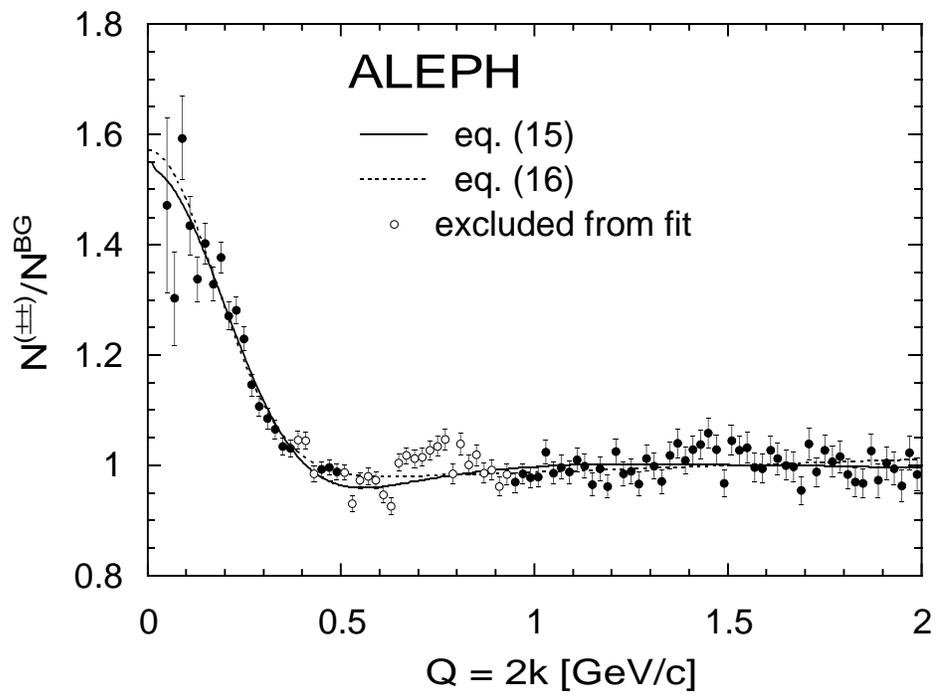
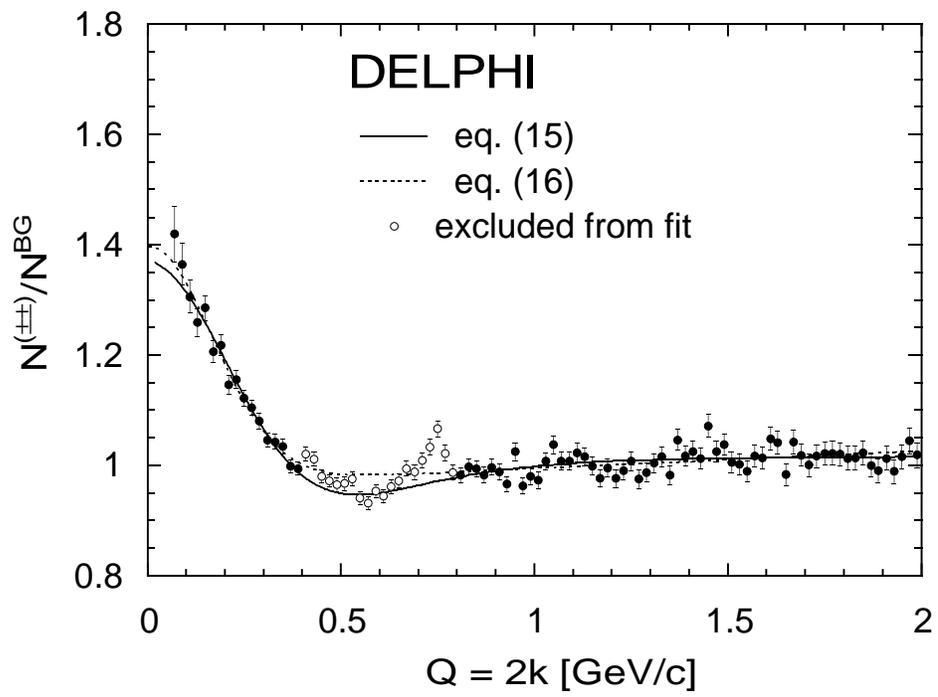
Fig. 3. $(a)$ Analysis of data of TPC Collaboration
\cite{aihara85} by (15) and (16). \\
$(b)$ The same as $(a)$ but for data of AMY Collaboration
\cite{amy95}. (The point at the smallest value of $Q$ is
neglected in analysis by means of (15).).\\
$(c)$ The same as $(a)$ but  data of OPAL Collaboration
\cite{opal93}.\\
$(d)$ The same as $(a)$ but  data of ALEPH Collaboration
\cite{aleph92}.\\
$(e)$ The same as $(a)$ but  data of DELPHI Collaboration
\cite{delphi92}. \\

\begin{center}
Table  Caption\\
\end{center}
Table I.  Analyses of data of the BEC by TPC, AMY, OPAL, ALEPH
and DELPHI  Collaborations; $a_0$=-1.00 GeV${}^{-1}$. The source size
parameters obtained by all collaborations are corrected  by $\beta =
\sqrt{2}R$.  There is no significant difference between those
corrected values (denoted by (*)) and values obtained in present
analysis. The AMY collaboration has used the fitting function $R_{\rm
mix}(Q)=C(1+f_{\rm \pi}(Q)\lambda \exp(-\beta^2 Q^2 /2))  (1+\gamma
Q),  f_{\rm \pi}(Q)=0.719-0.070Q+0.056Q^2-0.020Q^3$.

\clearpage
\pagestyle{empty}
\setlength{\oddsidemargin}{-1cm}
%%%%%%%%%%%%%%%%%%%%%%%%%%%%%%%%%%%%%%%%%%%%%%%%%%%%%%%%%%
%% ##################################################
%%  table  table  table  table  table  table  table
\begin{table}[t]
\begin{center}
\begin{tabular}{crrrrrr}
\hline
              &      & $\beta$~[fm]         & $\lambda$
              & $\gamma$         & $c$             & $\chi^2/$~NDF \\
\hline\hline
TPC           &&&&&& \\
$[20]$:       &      & $0.92 \pm0.06^\ast$  & $0.61\pm0.05$
              & $-$~~~~~~~       & $-$~~~~~~~      & $-$~~~~       \\
Our analyses: & (15) & $0.737\pm0.050$      & $1.097\pm0.042$
              & $-0.000\pm0.020$ & $1.002\pm0.022$ & $44.2/35$     \\
              & (16) & $0.912\pm0.062$      & $0.611\pm0.054$
              & $0.083\pm0.025$  & $0.881\pm0.023$ & $41.0/35$     \\
\hline
AMY           &&&&&& \\
$[21]$:       &      & $0.823\pm0.088^\ast$ & $0.392\pm0.041$
              & $0.033\pm0.041$  & $0.935\pm0.016$ & $90.2/93$     \\
Our analyses: & (15) & $0.460\pm0.039$      & $0.947\pm0.033$
              & $-0.078\pm0.012$ & $1.203\pm0.030$ & $106.8/95$    \\
              & (16) & $0.854\pm0.088$      & $0.286\pm0.030$
              & $0.030\pm0.013$  & $0.956\pm0.016$ & $94.1/96$     \\
\hline
OPAL          &&&&&& \\
$[22]$:       &      & $1.124\pm0.021^\ast$ & $0.846\pm0.025$
              & $-$~~~~~~~       & $-$~~~~~~~      & $336/73$      \\
Our analyses: & (15) & $1.090\pm0.035$      & $1.043\pm0.025$
              & $0.003\pm0.004$  & $0.991\pm0.005$ & $124.4/74$    \\
              & (16) & $1.339\pm0.035$      & $0.713\pm0.036$
              & $0.040\pm0.004$  & $0.936\pm0.004$ & $118.7/74$    \\
\hline
ALEPH         &&&&&& \\
$[23]$:       &      & $1.14\pm0.06^\ast$   & $0.48\pm0.03$
              & $0.02\pm0.01$    & $0.97\pm0.01$   & $77/70$       \\
Our analyses: & (15) & $0.917\pm0.037$      & $1.070\pm0.024$
              & $-0.016\pm0.008$ & $1.032\pm0.011$ & $89.0/69$     \\
              & (16) & $1.128\pm0.037$      & $0.630\pm0.030$
              & $0.024\pm0.009$  & $0.964\pm0.011$ & $87.3/69$     \\
\hline
DELPHI        &&&&&& \\
$[24]$:       &      & $1.16\pm0.04^\ast$   & $0.45\pm0.02$
              & $-$~~~~~~~       & $-$~~~~~~~      & $89/73$       \\
Our analyses: & (15) & $0.871\pm0.027$      & $0.946\pm0.016$
              & $-0.001\pm0.006$ & $1.020\pm0.008$ & $90.3/73$     \\
              & (16) & $1.170\pm0.039$      & $0.451\pm0.020$
              & $0.033\pm0.007$  & $0.963\pm0.008$ & $89.1/73$     \\
\hline
\end{tabular}
{\Large
\vspace*{1.5cm}

Table I
}
\end{center}
\end{table}
%%  table  table  table  table  table  table  table
%% ##################################################
\end{document}